\PassOptionsToPackage{table}{xcolor}
\documentclass[journal]{vgtc}                     % final (journal style)
\onlineid{3095}

\vgtccategory{Research}

\vgtcpapertype{algorithm/technique}
\newcommand{\us}[1]{Volume Encoding Gaussians}
\newcommand{\uss}[1]{VEG}

\usepackage{amssymb}
\usepackage{amsmath}
\usepackage{graphicx}
\usepackage{booktabs}
\usepackage{multirow}

\newcommand{\new}[1]{#1}

\title{Volume Encoding Gaussians: Transfer Function-Agnostic 3D Gaussians for Volume Rendering}

\author{%
  Landon Dyken,
  Andres Sewell, 
  Will Usher,
  Nathan Debardeleben,
  Steve Petruzza,
  Sidharth Kumar
}

\authorfooter{
  \item
  	Landon Dyken and Sidharth Kumar are with the University of Illinois Chicago.
  	E-mail: \{ldyke | sidharth\}@uic.edu.
  \item
  	Andres Sewell and Steve Petruzza are with Utah State University.
  	E-mail: \{a02024444 | steve.petruzza\}@usu.edu.
  \item 
    Will Usher is with Luminary Cloud.
  	E-mail: will@willusher.io.
  \item 
    Nathan Debardeleben is with Los Alamos National Laboratory.
    E-mail: ndebard@lanl.gov.
}

\abstract{%
Visualizing the large-scale datasets output by HPC resources presents a difficult challenge, as the memory and compute power required become prohibitively expensive for end user systems. Novel view synthesis techniques can address this by producing a small, interactive model of the data, requiring only a set of training images to learn from. While these models allow accessible visualization of large data and complex scenes, they do not provide the interactions needed for scientific volumes, as they do not support interactive selection of transfer functions and lighting parameters. To address this, we introduce \textit{\us{} (\uss{})}, a 3D Gaussian-based representation for volume visualization that supports arbitrary color and opacity mappings. Unlike prior 3D Gaussian Splatting (3DGS) methods that store color and opacity for each Gaussian, \uss{} decouple the visual appearance from the data representation by encoding only scalar values, enabling transfer function-agnostic rendering of 3DGS models. To ensure complete scalar field coverage, we introduce an opacity-guided training strategy, using differentiable rendering with multiple transfer functions to optimize our data representation. This allows \uss{} to preserve fine features across a dataset's full scalar range while remaining independent of any specific transfer function. Across a diverse set of volume datasets, we demonstrate that our method outperforms the state-of-the-art on transfer functions unseen during training, while requiring a fraction of the memory and training time.
}

\keywords{Volume Rendering, 3D Gaussian Splatting}

\renewcommand{\manuscriptnotetxt}{}

\graphicspath{{figs/}{figures/}{pictures/}{images/}{./}} % where to search for the images

\usepackage{mathptmx}      % use matching math font

\begin{document}

%%%%%%%%%%%%%%%%%%%%%%%%%%%%%%%%%%%%%%%%%%%%%%%%%%%%%%%%%%%%%%%%
%%%%%%%%%%%%%%%%%%%%%% START OF THE PAPER %%%%%%%%%%%%%%%%%%%%%%
%%%%%%%%%%%%%%%%%%%%%%%%%%%%%%%%%%%%%%%%%%%%%%%%%%%%%%%%%%%%%%%%

\firstsection{Introduction}

\maketitle

\label{sec:introduction}
%\will{What's the problem and why's it important?}
%\will{What's state of the art and why's it not good enough?}
%\will{What do we do about it that's better?}
%\will{Now talk about our great contributions}
%\will{Our new thing is amazing because: A, B, C.
%Here's a bulleted list of contributions to make it super obvious:}

% \sid{The introduction feels a bit underwhelming; maybe we need to add a few sentences explicitly saying why we are better than our competitors... and what are}

As high-performance computing (HPC) systems continue to grow in scale, scientific simulations are generating
ever-larger and more complex datasets. These simulations frequently use adaptive mesh refinement (AMR)~\cite{anderson2024fun3d, flash_2014, lava_2016,BERGER198964, arndtDealIIFiniteElement2021, bursteddeP4estScalableAlgorithms2011} or unstructured meshes~\cite{anderson2024fun3d, economonSU2OpenSourceSuite2016, offermansAdaptiveMeshRefinement2020, palaciosStanfordUniversityUnstructured2013} to reduce the memory footprint of output volumes. However, visualizing these volumes remains a significant challenge: datasets often exceed GPU memory, and visualization itself may require HPC resources, limiting interactivity and accessibility.

To address these challenges, recent work has explored machine learning (ML) to enable large-scale volume visualization, both for direct compression of volumetric datasets~\cite{Weiss_2022, wuInteractiveVolumeVisualization2023, wu2024interactivevolumevisualization, lu_neurcomp_2021} and for reducing the compute requirements during rendering~\cite{weiss_volumetric_2021, weiss_adaptive_2022, dyken_isosurface_2024, bauer_fovolnet_2023}. At the same time, novel view synthesis (NVS) techniques have emerged as a powerful method for representing 3D scenes within small memory and compute budgets. Requiring only a sparse set of training images depicting the original scene, a NVS model can learn to infer viewpoints at unseen angles, allowing for interactive rendering of a dataset with a memory footprint independent of the original size. Additionally, image-based training allows NVS techniques to natively support arbitrary data formats, as long as they can be rendered to a set of images. This presents an opportunity for volume visualization, where existing ML techniques have largely focused on structured volumes rather than AMR or unstructured data. 

Within the field of NVS, 3D Gaussian Splatting (3DGS) has shown itself as an impressive technique for real-time rendering, and has enjoyed an explosion of follow-up work since its initial debut~\cite{3dgs, 3dgssurvey}. In 3DGS, a scene is represented by a set of 3D Gaussians, each defined by a position, scale, rotation, view-dependent color, and opacity. These Gaussians are projected into screen space and alpha-blended to form the final image. Each Gaussian's parameters are optimized by an ML framework using differentiable rendering, and the training pipeline dynamically densifies and prunes the total number of Gaussians. Once optimized, 3DGS enables fast and high-quality rendering of 3D scenes. 
However, base 3DGS is fundamentally designed for surface-based, photogrammetry-derived datasets, not scientific volumes, as it embeds view-dependent color and opacity into each Gaussian. This makes 3DGS incompatible with the workflows that are typical for volume visualization, where users must be able to adjust color and opacity mappings interactively by changing transfer functions (TFs). To address this, previous work has extended 3DGS for volume visualization tasks~\cite{ivrgs_2025, nli4volvis_2025} using composition of multiple sets of \textit{editable Gaussians} that allow for adjusting color, opacity, and lighting after training. Each set of editable Gaussians is trained using a collection of multi-view images rendered from a single TF. By using many TFs that highlight disjoint areas of the volume's range, the composed set can represent the original data. While this method is accurate for the TFs the model was trained on, it is unable to robustly render unseen TFs. In addition, because the method involves composing multiple sets of 3DGS models to produce a full model, its training time and memory footprint quickly grow as more sets are required to represent larger datasets.

To address these issues, we present \textit{\us{} (\uss{})}, a novel 3DGS representation tailored for scientific volume visualization. Rather than storing color and opacity for each Gaussian, \uss{} store only scalar field values, decoupling the data representation from its visual properties. This allows \uss{} to support fully transfer function-agnostic rendering, where arbitrary color and opacity mappings can be applied dynamically at render time. Additionally, the entire function range of a volume can be accurately represented with only one set of \uss{}, greatly reducing memory and training time compared to methods which rely on composing multiple models. 
% The result is a high-quality representation of the original volume dataset that preserves spatial features while offering real-time volume rendering performance. 

% To build a \uss{} representation, we introduce a direct initialization strategy that avoids the need for prior scene reconstruction and instead instantiates Gaussians directly from structured and unstructured volume data. Training of \uss{} parameters then proceeds through differentiable rendering using ground truth images of the dataset. To allow the \uss{} representation to learn the complete function range of the data, we \will{this is a new thing right? so instead of "use" we should say "introduce a new method for opacity-guided..."} use opacity-guided training with multiple transfer functions that expose different internal structures of the volume. An overview of our method can be seen in~\Cref{fig:pipeline}. 
% \will{rework sentence to be more impactful. How compressed, what is "flexible"}

%We envision this being used as a high-quality approximation of large simulation data, which takes up less disk space while also being more efficient to render. This could potentially enable datasets to be rendered locally that would otherwise require HPC resources to visualize.We believe that this work paves the way for more accessible and efficient volume rendering, enabling high-quality visualization of large simulation data without the need for extensive computing resources.
Our work presents the following contributions: 
\begin{itemize}
    \item A method for novel view synthesis of volume data that can faithfully visualize color and opacity mappings unseen during training by approximating the underlying scalar field.
    \item An opacity-guided training technique that allows representing the full function range of volume data without separate training and composition for multiple transfer functions.
    \item An evaluation demonstrating that \uss{} achieve higher reconstruction quality on unseen transfer functions than the state-of-the-art, while having greatly improved training time and memory footprint (average $6.8\times$ faster training time and $17.0\times$ smaller files).
    % \item \will{We evaluate our work against ... and demonstrate it's awesome...}
\end{itemize}
\section{Related Work}
\label{sec:related_work}
% \will{too many sentences start with Because}
Interactive rendering of large-scale volumes has long been a focus area in scientific visualization~\cite{bunyk_simple_1997,weiler_hardware-based_2003,silva_survey_2005,sartonStateoftheartLargeScaleVolume2023}. Ever larger data sets and the desire for faster rendering and higher visual quality pose a continuous
challenge, while new hardware capabilities provide opportunities for new techniques~\cite{rathke_simd_2015,morrical_efcient_2019,morrical_accelerating_2020,morrical_quickclusters_2023,sahistanRaytracedShellTraversal2021}. For a current overview of the field, we refer the reader to the survey by Sarton et al.~\cite{sartonStateoftheartLargeScaleVolume2023} for more details.

In this work, we propose an image-based machine learning method for volume visualization. While our work does not produce a neural representation, it is similar to methods that do in that we use machine learning to optimize our \uss{} to represent volume data. Because of this, we review neural representations for compressing and rendering volume data in Section~\ref{sec:related_work-neural}. Next, we review image-based methods for volume visualization in Section~\ref{sec:related_work-image}. Finally, since our work draws heavily from 3DGS, we give background for this work and related methods in Section~\ref{sec:related_work-3dgs}.

\subsection{Neural Representations for Volume Data}
\label{sec:related_work-neural}
Neural representations for volume data, also known as scene representation networks (SRNs), implicitly model a volumetric function from input positions to scalar values using neural networks. Because neural networks can be sampled quickly and represent data without scaling linearly with dataset size, they offer a compressed representation that can be used for rendering. While there has been much work in this field (see the recent survey by Wang and Han~\cite{wang_dl4scivis_2023}), here we will focus on papers that apply to scientific volume data. Weiss et al.~\cite{weiss_differentiable} showed the possibility of reconstructing volume data from ground truth images using differentiable volume rendering. Jain et al.~\cite{griffin_compressedvr_2017} presented a neural method for compressed volume rendering of multivariate time-varying volumes. Kim et al. created NeuralVDB~\cite{kim_neuralvdb_2024}, which used neural networks to further improve the compression ability of the VDB~\cite{museth_vdb_2013} data structure for sparse volumes. 
% There is a recent trend in volume visualization for state-of-the-art methods to adapt groundbreaking work in computer vision to representing volume data, with large success. 
Lu et al. created \textit{Neurcomp}~\cite{lu_neurcomp_2021}, which  adapted SIREN's~\cite{sitzmann_siren_2020} implicit neural representation for efficient volume data compression. Weiss et al. developed fV-SRN~\cite{Weiss_2022}, which used dense-grid encoding~\cite{takihawa_dense_2021} and fast tensor core inference to build a neural representation that supports interactive rendering. Wu et al. then improved on fV-SRN's performance (in terms of reconstruction quality, training time, and rendering speed) with InstantVNR~\cite{wu2024interactivevolumevisualization}, which utilizes multi-resolution hash encoding~\cite{muller_multi_2022}, along with out-of-core training and batch inference during rendering. 

Although our method is similar to these representations in its goal of enabling visualization of large-scale volumes, there are multiple advantages arising from our method being image-based. First, image-based methods easily fit into simulation workflows, where output images can be written at each timestep and later used for training without transfer of the prohibitively large output volumes. Next, image-based models naturally support any data type, while only a small set of neural representations have been demonstrated to work directly with AMR or unstructured volume data~\cite{liu24_uginr, son2025_mcinr}. Finally, image-based methods offer performance that is completely decoupled from the original data size, only depending on image resolution and scene complexity. This allows training of image-based models to be done even on consumer hardware, rather than requiring HPC resources. 
% We hope to continue this trend by adapting the use of 3D Gaussians to represent volume data. Importantly, our work is the first to specifically address unstructured volumes, as the related work has so far focused solely on building representations for structured data. In addition, we uniquely explore the possibility of learning an explicit point-based representation for building a model of volume data. 
\subsection{Image-Based Methods for Volume Visualization}
\label{sec:related_work-image}
As image-based rendering and modeling methods are not tied to dataset size, they are important in enabling volume visualization for use cases where traditional rendering methods have prohibitively high compute and memory requirements. Early work in this area from Tikhonova et al.~\cite{tikhonova_volume_2010, tikhonova_time_2010} involved creating exploratory images, which allowed for changing color and opacity mappings of a rendered image without accessing the original volume data. Since then, image-based work has been extended to support synthesis of images from unseen view points using various ML approaches. Berger et al.~\cite{berger_generative_2019} proposed pre-trained generative adversarial networks to synthesize volume-rendered images and explore the space of transfer functions while guiding user selection. \new{Yao et al.~\cite{yao_revolve_2025} used a NeRF-based approach to train a network to represent pre-shaded volumetric scenes from a small number of rendered images. VisNerf~\cite{yao_visnerf_2025} presented a multidimensional radiance field representation that allows for novel  view synthesis of volumetric scenes with interactive control over transfer functions, isovalues, timesteps, and simulation parameters.} Most importantly to our work, iVR-GS~\cite{ivrgs_2025} showed how 3D Gaussian splatting can be extended for volume visualization by implementing Blinn-Phong lighting and editable Gaussian colors after training. Instead of directly storing color for each Gaussian, they store a palette color common to a set of Gaussians, train an offset color for each Gaussian, then produce each Gaussian's final color during rendering by adding the offset color to the palette color. This allows them to edit color after training by modifying the palette color of the set. We differ from iVR-GS in that we fully remove color and opacity attributes for each Gaussian, and instead allow our model to learn the scalar field represented by volume data directly. In doing so, we remove the need to train multiple models for each desired TF, achieving a transfer-function independent model after training.  

\subsection{3D Gaussian Splatting}
\label{sec:related_work-3dgs}
Splatting methods involve projecting a set of 3-dimensional primitives directly onto a 2-dimensional image, then $\alpha$-blending the primitives in order of their depth. Zwicker et al.~\cite{zwicker_splats_2001} first proposed the use of Gaussian filters as a primitive for splatting, showing their effectiveness for rendering both surface and volume data. Since the landmark work on utilizing neural radiance fields (NeRF)~\cite{mildenhall_nerf_2021}, novel view synthesis has seen an explosion of work. While neural radiance fields have been shown to produce highly accurate renderings, they regularly suffer from high training times and low rendering frame rates, reducing their applicability for real-time interaction. Kerbl and Kopanas et al.~\cite{3dgs} noted that, while the rendering algorithm for Gaussian splatting is very different, its image formation model is similar to traditional volume rendering methods such as those in neural radiance field rendering (NeRF)~\cite{mildenhall_nerf_2021}. At the same time, the explicit nature of 3D Gaussians enables fast rendering compared to the expensive sampling necessary for rendering with the continuous implicit neural representation of NeRFs. 3D Gaussian splatting~\cite{3dgs} was developed to optimize an explicit representation for novel view synthesis, allowing for greatly improved rendering performance while maintaining reconstruction quality.

Originally developed for photogrammetry-based scene reconstruction, in 3DGS each Gaussian stores an opacity along with view-dependent color using spherical harmonics, similar to previous work~\cite{yu_and_fridovichkeil2021plenoxels, muller_multi_2022}. In addition, initializing a set of 3D Gaussians relies on point clouds derived from NeRFs, COLMAP~\cite{schoenberger2016mvs, schoenberger2016sfm} or similar structure-from-motion pipelines. These features of 3DGS are inherently designed for real-world image datasets, making them unsuitable for scientific volume data, where direct volume rendering~\cite{Max1995OpticalModelsDirect} techniques are commonly used to map scalar field values to optical properties using transfer functions consisting of color and opacity maps. Arbitrary transfer functions are needed for data exploration, so volume rendering must support interactive transfer function adjustments, meaning that hard-coded color and opacities in the data representation are not suitable. \new{While recent work~\cite{han_isosurface_2025} has used distributed 3DGS~\cite{zhao2024scaling3dgaussiansplatting} for visualization of large scientific datasets, this method supports only a single static transfer function per 3DGS model.} In our work, we remove this limitation through our development of \us{} which represent the underlying data field of a volume directly. Moreover, we replace the use of scene reconstruction for initialization, instead using the given volume dataset or random initialization.

Our work also takes advantage of recent extensions to 3DGS with regards to lighting effects and reduced memory footprint. Gao et al.~\cite{gao_relightable_2024} introduced relightable Gaussians, which augmented 3DGS with extra parameters to support Bidirectional Reflectance Distribution Function (BRDF) lighting. iVR-GS~\cite{ivrgs_2025} adapted this method to the Blinn-Phong reflection model, which is more common to volume rendering use cases. Our method adopts their technique of Blinn-Phong lighting for producing shaded images with our own model. The original 3DGS formulation had an unnecessarily high memory requirement, and several works have been created to address this limitation~\cite{papantonakis_compressedgs_2024, niedermayr_compressedgs_2024, fan_lightgaussian_2025}. We adopt these methods by utilizing vector quantization to compress our models while maintaining reconstruction quality, clustering most model parameters into a codebook for storage.
\section{Design of \uss{}}
\label{sec:design}
% \begin{figure}[t]
%     \centering
%     \includegraphics[width=\linewidth]{figs/gaussian-volumes.pdf}
%     \caption{Visualization of sampling for structured (left) and unstructured (right) volumes. Structured volumes define scalar fields over a regular grid, while unstructured volumes must define an explicit geometry.}
%     \label{fig:volumes}
%     \vspace{-1.5em}
% \end{figure}

In this section, we give background on volume data types (Section~\ref{sec:types}), then describe the \uss{} representation (Section~\ref{sec:representation}), an extension of 3DGS tailored to scientific volume visualization. Finally, we present how \uss{} are used for differentiable rendering with arbitrary transfer functions (Section~\ref{sec:rendering}).  

\subsection{Volume Types}
\label{sec:types}
A structured volume is defined by a set of scalar field values on a regular grid, with neighboring points assumed to be connected grid cells. The geometry of a structured grid is defined implicitly; only the data values need to be stored along with the grid dimensions and spacing. Any point in space can then be interpolated directly from the surrounding grid points. In contrast, unstructured volume data requires explicit storage of the geometry over which the scalar field is defined. Given a position in space, a test is performed to determine what cell in the volume, if any, contains it, before interpolating between that cell's vertices to find the scalar value at that position. 
% An example of sampling volume data is shown in Figure ~\ref{fig:volumes}. 

% Both structured and unstructured volume data can be defined as functions that map positions in a subset of 3-dimensional space to values in a scalar field, i.e. $\Phi: \Omega \subset \mathbb{R}^3 \rightarrow \mathbb{R}, (x,y,z)\mapsto \Phi(x,y,z)=v
% $. They differ in that, while the domain of a structured volumetric function is easily defined in space by its origin and bounds, the domain of an unstructured volumetric function has no set structure; the function can be defined over any region matching the unstructured volume's geometry. In addition, the complexity of this function is not uniform with respect to spatial area, as values in an unstructured volume can have arbitrary spacing. 
% In practice, efficient representation of scalar fields with massive differences in spatial feature density is the main reason for using unstructured volumes. 

% Current neural representations for structured volume data~\cite{lu_neurcomp_2021,Weiss_2022,wu2024interactivevolumevisualization} utilize scene representation networks~\cite{park_continuous_2019,Mescheder_recon_2019,nerf}, which implicitly model a volumetric function from input positions to scalar values using neural networks. Because of the unique features of unstructured volumetric functions mentioned above, the extension of neural methods to unstructured volume data is not readily apparent. 
\subsection{The \uss{} Representation}
\label{sec:representation}
In our work, we propose an explicit, point-based representation to encode a volume's geometry and underlying scalar field. To do this, we use a 3D Gaussian formulation, where each point is defined as an ellipsoid centered at position $\mu$ with 3-dimensional scaling and rotation matrices, $S$ and $R$. This allows the points to translate, stretch, and rotate independently in each dimension, giving the representation freedom to adapt to the arbitrary geometries of structured and unstructured volumes. Each Gaussian stores a value, allowing it to record the scalar field over the covered area of the volume's geometry. In this way, our \us{} (\uss{}) give a compact representation that can encode the full information of the volume being rendered without storing any connectivity information for unstructured volumes.
% While unstructured volumes are made up of both scalar-valued points and the connectivity information describing the cells connecting those points, the geometry-capturing nature of 3D Gaussians allows our representation to be simply made up of scalar-valued Gaussians, without storing any connectivity information. 

\begin{figure*}[t]
    \centering
    \includegraphics[width=\linewidth]{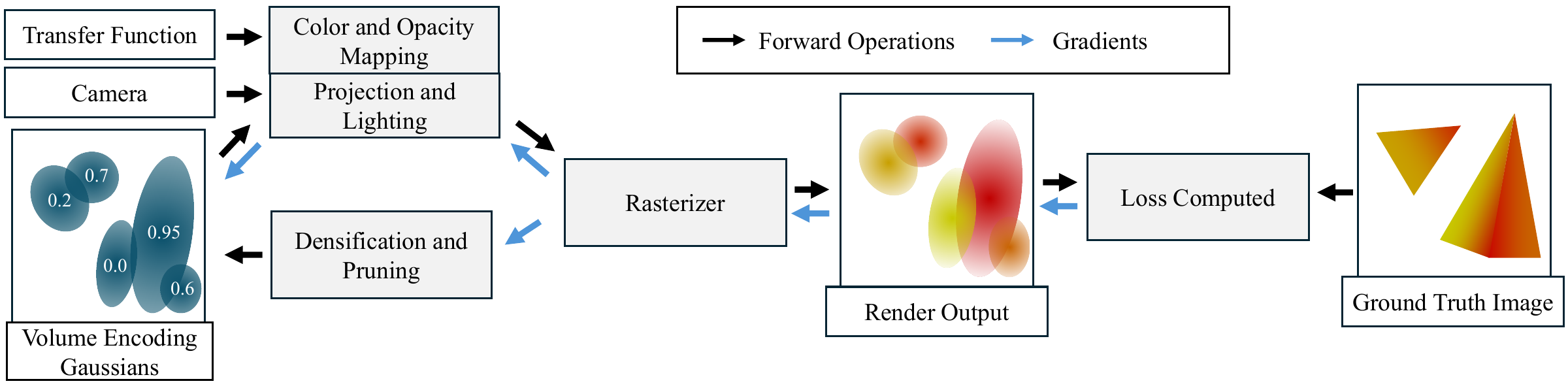}
    \vspace{-1em}
    \caption{Diagram of our training pipeline, where operation flow is depicted with black arrows and gradient backpropagation is depicted with blue. We conduct image-space training, in which our differentiable rendering algorithm is used to optimize \us{} from loss computed against ground truth images. Each training step, we render our \uss{} by opacity/color mapping with a given transfer function, applying lighting effects with the Blinn-Phong model, projection into 2D with a given camera view, then rasterization and blending of the 2D colored primitives. We then compare the rendered output against shaded reference images created from volume rendering with an emission-absorption model. After computing loss, gradients are backpropagated through to the scaling, rotation, position, scalar values, weight and lighting parameters of the \uss{} for optimization. Gradients are also used for densification and pruning of the \uss{}, in which the density of 3D Gaussians in space is adaptively controlled. }
    \label{fig:pipeline}
    \vspace{-1em}
\end{figure*}

\subsection{Rendering of \uss{}}
\label{sec:rendering}
The goal of our representation is to approximate  direct volume rendering~\cite{Max1995OpticalModelsDirect} of the ground truth dataset. Specifically, we target raymarching with the emission-absorption model, where the color of each pixel is given by accumulating the contribution of $N$ samples along the ray using $\alpha$-blending with the discretized equation:

\begin{equation}
P = \sum_{i=0}^{N} c_i \cdot \alpha_i \cdot \prod_{j=0}^{i-1} (1 - \alpha_j)
\end{equation}

where samples $i$ of the represented scalar field produce each opacity $\alpha_i$ (emission) and color $c_i$ (absorption) through transfer function application and lighting effects. We use 3D Gaussian splatting to model Equation 1 approximately, where the \us{} representing a scalar field are sampled through splatting. To render each Gaussian, the scaling and rotation matrices are used to create the Gaussians' 3-dimensional covariance matrix, which is then projected to a 2-dimensional covariance matrix using the method of Zwicker et al.~\cite{zwicker_splats_2001}. This converts the 3D Gaussian in object space into a 2D Gaussian in image space. The list of 2D Gaussians intersecting the screen can then be used to create the rendered image using the efficient tile-based rasterizer of Kerbl and Kopanas et al.~\cite{3dgs}. This rasterizer employs $\alpha$-blending of splatted Gaussians in front-to-back order, where a pixel accumulates each Gaussian intersecting it by evaluating its 2D covariance matrix at the pixel's position (using the method of Yifan et al.~\cite{yifan_surface_2019}). This solves Equation 1, effectively approximating the emission-absorption model for raymarching-based volume rendering. 

To determine how each \uss{} contributes to the color and opacity of a pixel, we first use a given transfer function to map each Gaussian to opacity and color. Rather than evaluating a transfer function when accumulating samples, it is more efficient to apply the transfer function as the first step of our rendering algorithm. This allows us to immediately cull Gaussians with values mapped to low opacities, saving computation time similarly to empty space skipping in traditional volume rendering. 

We found that, although this transfer function mapping follows the principles of volume rendering, it leads to poor training performance due to limiting the model's ability to optimize opacity directly. This is important, as 3DGS effectively uses opacity to weigh the importance of different Gaussians, increasing the visibility of those that are useful and removing those with low opacity for density control. To reintroduce this ability into \uss{}, we augment our Gaussians with a trainable weight parameter analogous to the original opacity, and output the final opacity for each \uss{} as the product of its transfer function-dependent opacity and its weight, i.e. $\alpha_i = O(v_i) \cdot w_i$ where $O(v)$ is the opacity mapping of the transfer function, $v_i$ is the Gaussian's scalar value and $w_i$ is the Gaussian's weight. 

Additionally, our model needs to support lighting effects, which are essential for 3D perception during rendering. For volume visualization, the most common lighting model used is Blinn-Phong. The Blinn-Phong model produces lighting effects by summing ambient ($c^a$), diffuse ($c^d$), and specular ($c^s$) colors output by samples, rather than just the transfer function-determined color $C(v)$. We adopt the technique of iVR-GS~\cite{ivrgs_2025} for Blinn-Phong lighting in our model by attaching ambient, diffuse, and specular coefficients ($k^a_i$, $k^d_i$, and $k^s_i$), a normal vector ($n_i$), and a shininess term ($\beta_i$) to each Gaussian. Then, given a light direction $l$, we produce the final color for each Gaussian as:

\begin{equation}
c_i = k^a_i \cdot C(v_i) + k^d_i \cdot C(v_i) \cdot |n_i\cdot l|+c^s_i, 
\end{equation}
\begin{equation}
c^s_i = \begin{cases}
k^s_i \cdot c_{white} \cdot |n_i \cdot \frac{v + l}{|v + l|}|^{\beta_i} , \text{if } |n \cdot l| > 0 \\
0, \text{otherwise}
\end{cases}
\end{equation}

This gives the final parameter list for each \uss{} ($G_i$) in our model as ($\mu_i$, $S_i$, $R_i$, $v_i$, $w_i$, $n_i$, $k^a_i$, $k^d_i$, $k^s_i$, $\beta_i$). In the backwards pass of rendering, gradients are computed manually for each Gaussian parameter using a modified version of Kerbl and Kopanas et al.'s~\cite{3dgs} rasterizer. In order to optimize the attached scalar values for each Gaussian, we backpropagate gradients with respect to color and opacity through the lighting equations and transfer function used to render that frame. We assume linear interpolation of color and opacity between each control point of a given transfer function, but allow for an arbitrary number of control points to be given. To constrain values within $[0,1)$ and ensure smooth gradients during training, we utilize a sigmoid activation function for storing scalar values and weights.

\begin{figure}[t]
    \centering
    \includegraphics[width=\linewidth]{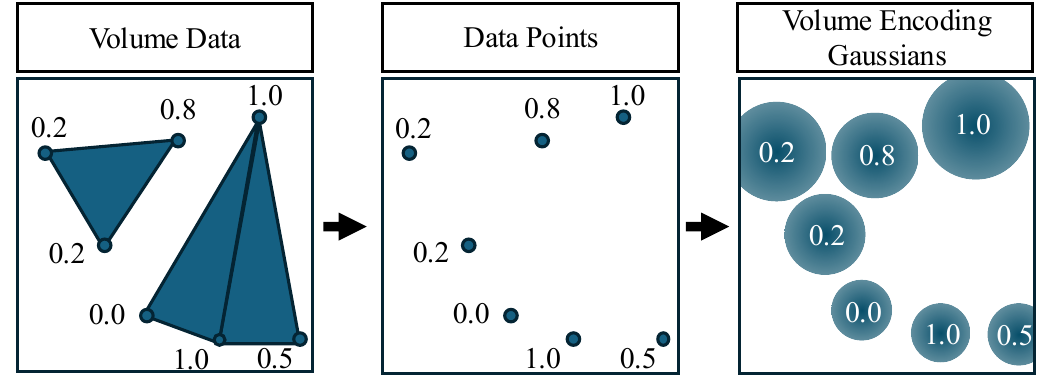}
    \caption{Diagram of our initialization strategy to create \uss{} from volume data. 
    % For both cell-centered and vertex-centered data, 
    Our approach begins by deleting all connectivity information, leaving only scalar-valued data points. Next, we extend these data points with 3D scaling and rotation matrices, so that each point is a 3D Gaussian with attached data value. These are initialized to make each initial 3D Gaussian a sphere with size relative to the density of other points around it, with lower density creating larger Gaussians.}
    \label{fig:initialization}
    \vspace{-1em}
\end{figure}

\section{Optimization of \uss{}}
\label{sec:optimization}
The essence of our method is our optimization of \uss{} to accurately represent volume data. In this section we outline our strategy, starting with how we initialize our models in Section~\ref{sec:initialization}. We then detail our method for training these models to completion in \Cref{sec:training}. 

\begin{figure*}[t]
    \centering
    \includegraphics[width=\linewidth]{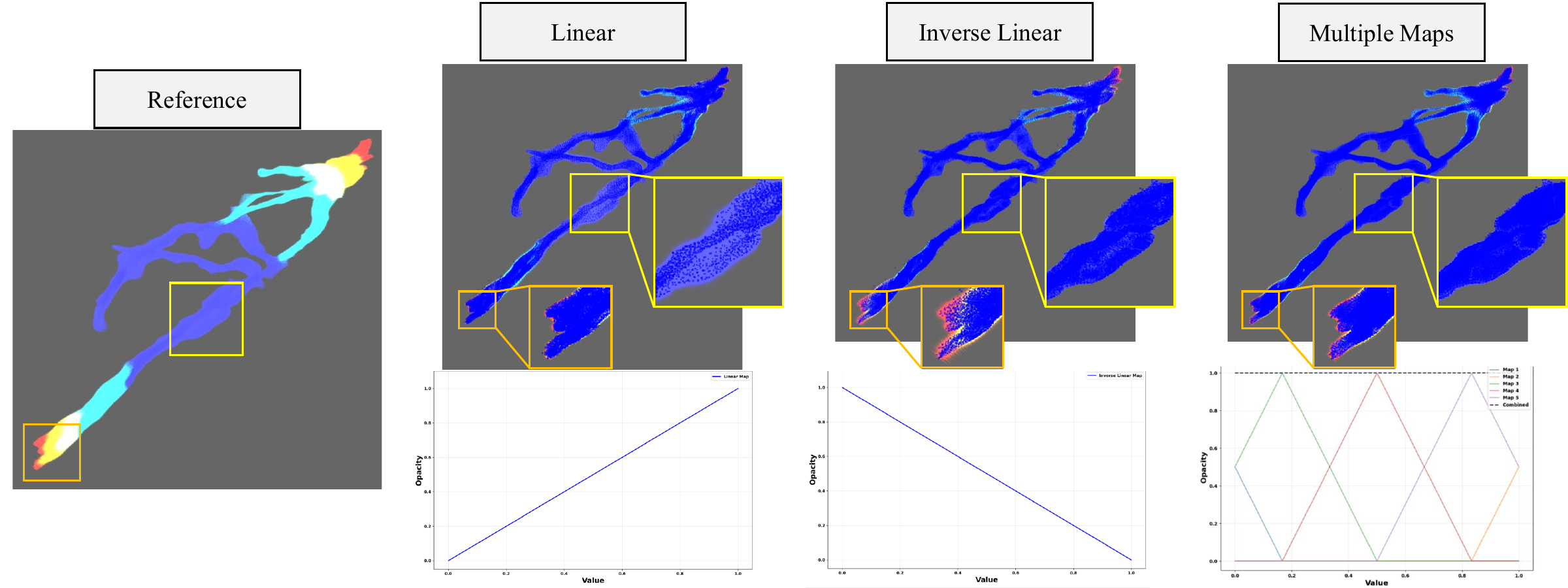}
    \caption{Gaussian density comparison of \uss{} models trained using either a linear opacity map, an inverse linear opacity map, or our multiple opacity map method using three steps on the RBL dataset. Reference shows volume rendering with a pale rainbow colormap, where high values are red and low values are pale blue. Gaussian centers are painted using deep blue for each of the trained models, on top of the reference image, using SuperSplat's\cite{supersplat} 3D Gaussian editor. Selected areas of high value are highlighted with an orange box, and areas of low value are highlighted with a yellow box. Notice training with the linear transfer function (which gives high opacity to high values) leads to greater Gaussian density in areas of high values and very low density in areas of low values, while the model trained with the inverse linear map does the opposite. Our multiple opacity map method leads to equal density for both areas of the function range.}
    \label{fig:opacity_maps}
    \vspace{-1em}
\end{figure*}

\subsection{Initialization}
\label{sec:initialization}
As shown in previous work~\cite{3dgs}, proper initialization can be essential for learning explicit scene representations. Because the input to our method is a volume dataset, it is important to initialize our \uss{} with some knowledge of the scalar field they will be trained to represent. To do this, we experimented with initializing \uss{} from either cell centers or vertices of a volume, but found both methods to give around equal performance, so present here only the vertex-centered method. 
% Most commonly, volume data is stored with values defining its scalar field attached to points in space, which are then connected as vertices of cells.
In order to initialize our \uss{}, we first drop the connectivity information and treat the scalar-valued vertices as a point cloud with each point storing a value. 
% In our cell-centered initialization, we first interpolate the value of the scalar field for the center of each cell in the volume, then again drop the connectivity information and obtain a point cloud of scalar-valued cell centers. From either cell-centered or vertex-centered point cloud, 
We then transform each point into a \uss{} centered at the same position with the same scalar value. We initialize each \uss{}'s scale depending on the proximity of its closest neighbors, with larger Gaussians in areas of empty space and smaller Gaussians in areas of high density. This process is shown in \Cref{fig:initialization}. 
% We found both methods of initialization to give around equal performance; in our evaluation, we use vertex-centered initialization. 
% We observe that the percentage of Gaussians dropped out here can significantly affect the final compression ratio and reconstruction quality, with the optimal value depending on data size and complexity. We explore how different dropout percentages affect training for different datasets in Section~\ref{sec:evaluation}. 

In practice, we randomly drop out all but a small number of data points before initializing the \uss{} \new{in order to create compact models, as each \uss{} can represent many data points.} This also makes our method easier to fit into HPC workflows, as only a small sampling of data is required to be stored for later training. For use cases where this is not possible, our method can also perform well with random initialization, as shown in Section~\ref{sec:ablation}. \new{The exact number of Gaussians needed for a \uss{} model depends on a dataset's complexity, and so we rely on adaptive density control during training (see Section~\ref{sec:training}) and only use a very rough estimate for initialization. For unstructured volumes, we also experimented with selecting data points uniformly in space rather than randomly from the unstructured mesh, but found this to have little effect on training; in our evaluation, we use random selection of data points for initialization.}

% \begin{figure}[ht]
%     \centering
%     \includegraphics[width=\linewidth]{figures/untrained.png}
%     \caption{Newly initialized Gaussian splat model, before training has occurred. (get a better picture than this)}
%     \label{fig:skull-untrained}
% \end{figure}

\subsection{Training}
\label{sec:training}
To optimize our \uss{}, we conduct image-space training using our fast differentiable rendering algorithm described in Section~\ref{sec:rendering}. The complete training pipeline is shown in Figure~\ref{fig:pipeline}. The model is iteratively optimized to match training images via backpropagation to each Gaussians' parameters. Through this, the \uss{} are able to learn both the geometry of the volume and value of the scalar field. 

In order to accurately recreate a dataset, the model must have training data that can allow it to learn all of the dataset's information. This process requires creating a robust set of training images with corresponding transfer functions and camera information. For our evaluation, we accomplish this by rendering a set of ground truth images using PyVista's~\cite{sullivan2019pyvista} VTK backend~\cite{vtkBook}. For these images, PyVista raymarches volumes with an emission-absorption model and shades them with Blinn-Phong lighting with a headlight from the camera. Images are rendered in a full camera orbit to ensure complete coverage of the given dataset.

To allow our model to render images for transfer functions never seen during training, we must ensure the \uss{} are able to reconstruct the full function range of the dataset. For this, the transfer functions used to create the training set of images must be carefully chosen. Because optimization is done from loss computed between images, transfer functions which highlight certain ranges of the scalar field effectively weight the set of \uss{} to reconstructing the regions of the dataset containing those ranges. We found that the mapping of values to color did not have large effect on the resulting trained \uss{}, as long as the colormap used was not diverging. Opacity mapping, on the other hand, was especially significant to deciding which areas of the volume were reconstructed. This is intuitive, since areas of value mapped to high opacity will be rendered with high opacity in resulting images, making them more impactful on loss, which leads optimization to assign more relevance towards reducing their error. 

% \steve{this next bit is unclear}
We take advantage of this to improve the reconstruction quality of our models with \textit{opacity-guided training}, where we use the opacity applied by transfer functions as an importance mapping to reconstruct all areas of the given dataset. To create our training set of images, we render each camera angle with multiple opacity maps highlighting different areas of the scalar field's range. We design these opacity maps so that, in combination, every area of the range is weighted as equally important. Because these opacity maps will be used for optimizing scalar values, it is also important that there is slope between each data point, in order for gradients to move values up or down during training. We build the set of opacity maps by first dividing the function range into a number of steps, which we take as an input parameter. We then create an opacity map centered in each step, with linear drop-off around this center having slope equal to the number of steps. A visual comparison of trained Gaussian density resulting from our opacity-guided training technique against naive approaches is given in \Cref{fig:opacity_maps}. \new{We evaluate the quality of models trained with our method against the naive approach, as well as investigate the number of opacity steps needed for accurate training in Section~\ref{sec:ablation}.}

% Depending on the scalar-value range of a given dataset, these opacity maps may yield entirely black images in the ground truth renders. For example, the Pig Heart dataset is mostly empty for about half of its function range, resulting in blank images for certain opacity maps highlighting this range. The inclusion of these blank images during training was found to significantly degrade the reconstruction quality of the \uss{} models. To mitigate this, each ground-truth image is evaluated and excluded from the training set when it is found to be too empty to contain enough information for training.

For each training iteration, a random camera angle and transfer function are chosen from the training set. The current set of \uss{} are then rendered and compared against the ground truth image for the same settings in order to compute loss. We compute image loss as a mixture of L1 and SSIM~\cite{wang_ssim_2004} loss, i.e. $(1-\lambda)L_1+\lambda L_{SSIM}$, where $\lambda$ weighs them against each other. We set $\lambda$ to 0.2 for all our tests, \new{borrowing this convention from other work in 3D Gaussian splatting~\cite{3dgs, ivrgs_2025}}. Additionally, we also optimize Gaussian normals directly using the normal consistency loss from Gao et al.~\cite{gao_relightable_2024}, and Gaussian lighting attributes ($k^a_i$, $k^d_i$, $k^s_i$, and $\beta_i$) using a small bilateral smoothness regularization term \cite{yao_neilf_2022, zhang_neilf_2023, ivrgs_2025}.

% This term takes the mean of the L2 norms for the inverted scaling of each Gaussian. This effectively penalizes Gaussians from becoming smaller, promoting them to represent the unstructured volume with fewer, larger Gaussians when possible to save memory and computation. The choice of $\lambda_{scaling}$ can affect compression factor and reconstruction quality, as increasing its value can reduce the number of Gaussians in the representation but also lead to under-reconstruction of high frequency features. We set our default value of $\lambda_{scaling}$ to $1e^{-5}$ based on initial tests and explore using different values in Section~\ref{sec:evaluation}. Additionally, we add another loss term $\mathcal{L}_{bounding}$ defined as:
% \begin{equation}
%     \frac{1}{N} \sum_{i=1}^{N} \sum_{d \in \{x,y,z\}} \left( \max(\mu_{i,d} - b_{d}^{\max}, 0) + \max(b_{d}^{\min} - \mu_{i,d}, 0) \right)^2
% \end{equation}
% where $b$ stores the bounding box (in terms of maximum and minimum x, y, and z) of the original volume being modeled. This loss helps to encourage the \uss{} to stay within the convex hull of the original dataset, preventing drift and maintaining consistency of coordinate position between the representation and original volume.

For best resulting models, training needs some way to adaptively control the density of Gaussians in space. This allows for removing Gaussians from areas where they are unneeded, saving memory and computation, and adding them to areas where the current number of Gaussians cannot accurately represent the volume's complexity. This is accomplished via densification and pruning heuristics during training. For both, we follow the method of Kerbl and Kopanas et al.~\cite{3dgs}. They show that, if a Gaussian has large view-space positional gradients, it is likely to be in a region that is not well reconstructed. Therefore, new Gaussians need to be added to that area, which is done by either cloning or splitting the Gaussian. For pruning, their work removes Gaussians with low attached opacity, so we follow this heuristic by using each \uss{}'s attached weight parameter $w_i$. We find these heuristics sufficient for creating accurately compact models depending on the complexity of the volume being reconstructed.

\section{Evaluation}
\label{sec:evaluation}

In this section, we perform an extensive evaluation of our method both quantitatively and qualitatively. We list our datasets and system setup in Section~\ref{sec:setup}. Our experimental study is to compare our method with the state-of-the-art image-based method for volume visualization, iVR-GS~\cite{ivrgs_2025}. We evaluate both methods with regards to reconstruction quality, memory footprint, training time, rendering performance, and generalizability to unseen transfer functions in Section~\ref{sec:benchmark}. \new{We then perform an ablation study of the design decisions of our method in Section~\ref{sec:ablation}}. Finally, we present and briefly describe a qualitative showcase of our method on more complex transfer functions in Section~\ref{sec:head}.

In addition, we also adapted the interactive renderer of Tang et al.~\cite{ivrgs_2025} to our use case, as presented in Figure~\ref{fig:renderer}. Please see the supplemental video for a demo of interaction with our renderer and comparison with iVR-GS. We additionally created a web application port of this renderer, accelerated with WebGPU\cite{webgpu_web}, to illustrate the potential of our method to enable lightweight client devices within a low memory footprint.

\begin{figure}[t]
    \centering
    \includegraphics[width=\columnwidth]{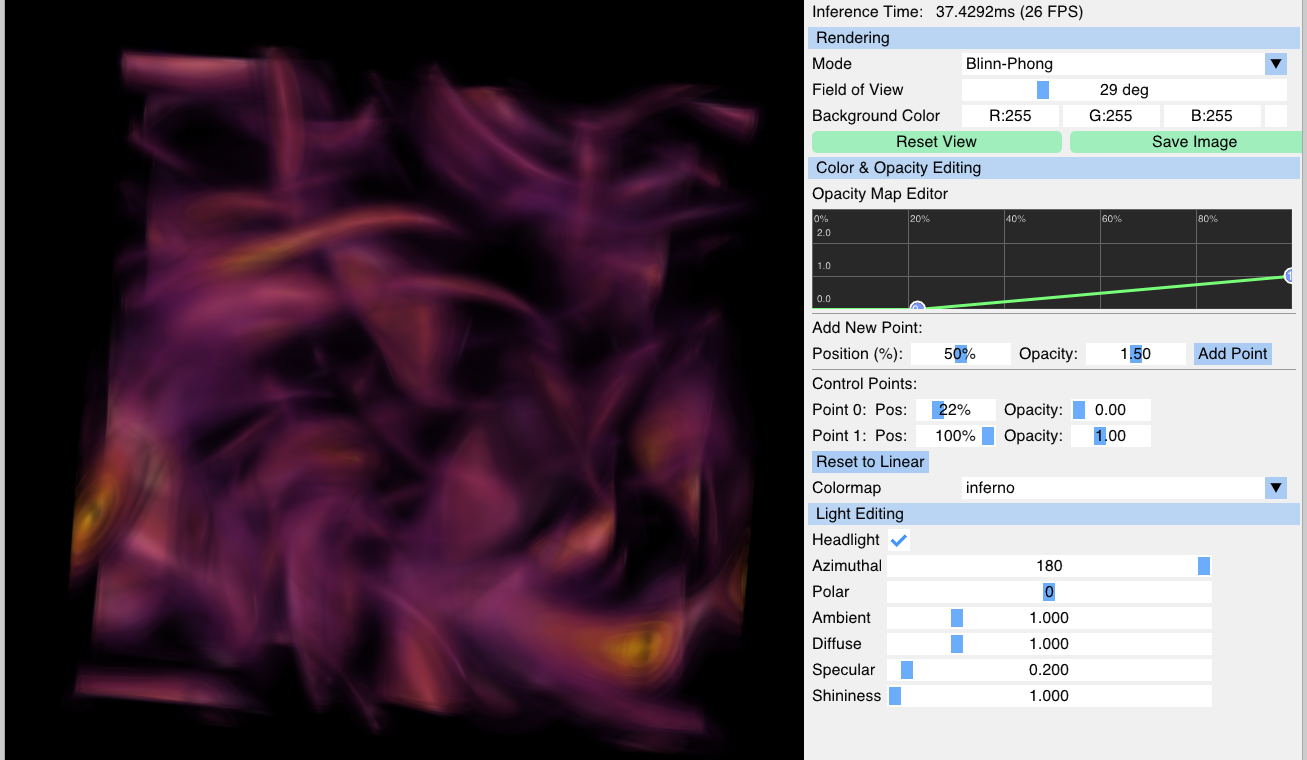}
    
    \caption{Interactive VEG renderer adapted from Tang et al.~\cite{ivrgs_2025}. Because \uss{} models are trained to generalize to arbitrary transfer functions, we replace their per-basic-set color and opacity sliders with selection of colormaps and an interactive opacity map creation tool. }
    \label{fig:renderer}
    \vspace{-1.5em}
\end{figure}

\begin{table}[h]
\centering
\begin{minipage}{0.48\columnwidth}
\centering
\begin{tabular}{@{}l|l@{}}
\toprule
Dataset & Dimensions \\ \midrule
Vortex & (512, 512, 512) \\
Chameleon & (1024, 1024, 1080) \\ \bottomrule
\end{tabular}
\end{minipage}%
\hfill
\begin{minipage}{0.48\columnwidth}
\centering
\begin{tabular}{@{}l|l@{}}
\toprule
Dataset & Tetrahedra \\ \midrule
RBL & 3.89M \\
Mito & 5.54M \\ \bottomrule
\end{tabular}
\end{minipage}
\caption{Chosen datasets and their size. Separated into structured volumes (left) and unstructured volumes (right).}
\label{tab:datasets}
\vspace{-1em}
\end{table}

\subsection{Setup}
\label{sec:setup}

\textbf{Datasets} To showcase the potential of \uss{} across different volume types, we benchmark our method on 4  datasets, consisting of 2 structured volumes (Vortex~\cite{ivrgs_2025} and Chameleon~\cite{scivisdata}) and 2 unstructured volumes (Mito and RBL)~\cite{rathke_simd_2015}. The size of each is given in Table~\ref{tab:datasets}.

\textbf{Systems} All training was performed on a workstation with an Nvidia A40 GPU and an AMD EPYC 9124 16-Core Processor. To illustrate the applicability of both methods to less powerful end-user systems, all rendering performance results were achieved on an XPS~17 laptop with an RTX~4070 Laptop GPU and an i9-13900H CPU.

% \textbf{Metrics} In order to evaluate our approach, we need to fairly compute metrics for reconstruction quality, compression, training time, and rendering performance. For reconstruction quality, we compare the output of our models to ground truth images on camera angles and transfer functions that were never used during training.  Each transfer function is used to render 16 camera angles, and average SSIM and PSNR values are computed between the ground truths and images created by our models. Finally, for rendering performance, we simply take the average time of the render calls used for the image quality benchmarks, removing the first 20 iterations to account for warmup time. 

\begin{table*}[t]
\centering
\resizebox{\textwidth}{!}{%
\begin{tabular}{@{}l|l|c|c|c|c|c|c|c|c|c|c|c|c@{}}
\toprule
\multirow{2}{*}{Dataset} & \multirow{2}{*}{Method} & \multicolumn{2}{c|}{Training TF} & \multicolumn{2}{c|}{Unseen Colormaps} & \multicolumn{2}{c|}{Broad Opacity} & \multicolumn{2}{c|}{Narrow Opacity} & Training & \# of & File Size & Avg. \\
& & PSNR & SSIM & PSNR & SSIM & PSNR & SSIM & PSNR & SSIM & Time (min) & Gaussians & (MB) & FPS\\
\hline
\multirow{2}{*}{Vortex} & iVR-GS & \textbf{37.83} & \textbf{0.99} & 28.41 & 0.96 & 23.20 & 0.92 & 21.69 & \textbf{0.90} & 219.8 & 3,848,171 & 90.2 & 3.9\\
& \uss{} & 35.19 & 0.98 & \textbf{33.98} & \textbf{0.97} & \textbf{25.85} & \textbf{0.92} & \textbf{22.22} & 0.89 & \textbf{29.7} & \textbf{216,328} & \textbf{4.9} & \textbf{40.9}\\
\hline
\multirow{2}{*}{Chameleon} & iVR-GS & \textbf{34.41} & \textbf{0.96} & 29.27 & 0.94 & 24.42 & 0.90 & \textbf{24.99} & \textbf{0.91} & 144.8 & 1,327,237 & 31.1 & 17.5\\
& \uss{} & 32.60 & 0.95 & \textbf{31.44} & \textbf{0.94} & \textbf{28.93} & \textbf{0.93} & 24.45 & 0.91 & \textbf{20.1} & \textbf{70,742} & \textbf{1.6} & \textbf{51.3}\\
\hline
\multirow{2}{*}{RBL} & iVR-GS & 37.38 & 0.99 & 34.03 & 0.99 & 27.83 & 0.98 & 23.49 & 0.98 & 49.8 & 195,384 & 4.6 & 47.1\\
& \uss{} & \textbf{39.83} & \textbf{0.99} & \textbf{40.65} & \textbf{0.99} & \textbf{32.81} & \textbf{0.99} & \textbf{31.08} & \textbf{0.99} & \textbf{11.3} & \textbf{30,218} & \textbf{0.7} & \textbf{61.2}\\
\hline
\multirow{2}{*}{Mito} & iVR-GS & \textbf{37.23} & \textbf{0.99} & 25.55 & 0.98 & 23.20 & 0.96 & 23.57 & 0.97 & 137.0 & 1,105,228 & 25.9 & 19.8\\
& \uss{} & 36.36 & 0.99 & \textbf{33.75} & \textbf{0.99} & \textbf{28.00} & \textbf{0.97} & \textbf{28.08} & \textbf{0.98} & \textbf{17.1} & \textbf{49,969} & \textbf{1.1} & \textbf{57.8}\\
\bottomrule
\end{tabular}
}
\caption{Evaluation metrics for novel view synthesis across different datasets for our method (VEG) and the state of the art for volume visualization (iVR-GS). Results show that our models have drastically improved training times and memory footprints. Additionally, our models outperform iVR-GS when testing on color and opacity mappings not seen during training (Unseen Colormaps, Broad Opacity, Narrow Opacity), although iVR-GS has greater reconstruction quality when testing on the transfer functions used for training. Best results for each metric are in bold.}
\label{tab:evaluation_results}
\vspace{-1em}
\end{table*}

\begin{figure*}[t]
    \centering
    \includegraphics[width=\linewidth]{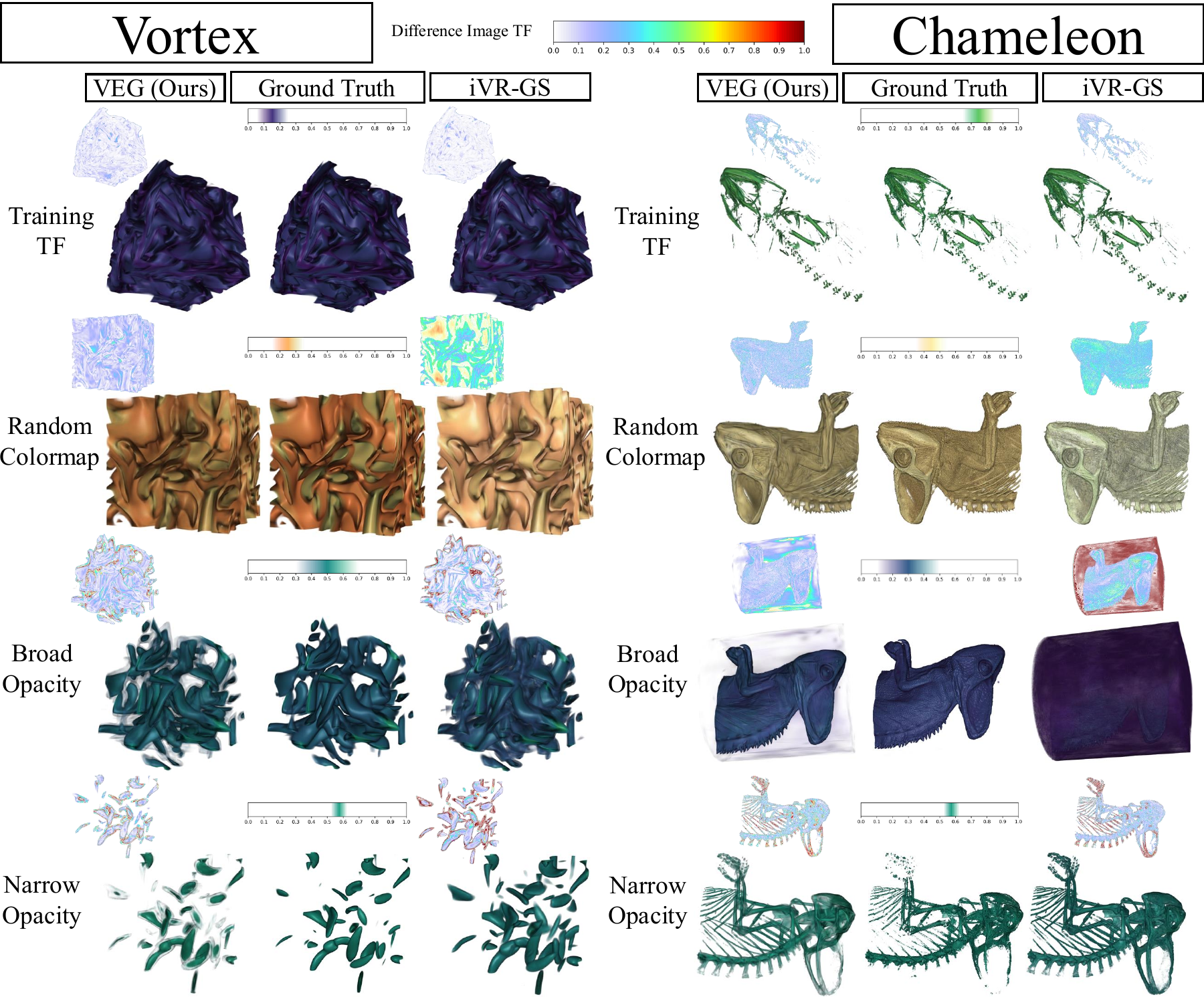}
    \caption{Qualitative comparison of our method with iVR-GS on the Vortex and Chameleon datasets. For each dataset and method, we show results from the tests run in Table~\ref{tab:evaluation_results}, \new{along with difference images showing perceivable difference in the CIE LUV color space between reconstructions and ground truth. Difference images use the Matplotlib 'jet' colormap with linear opacity map, shown at top, to highlight areas of low error (purple to blue), medium error (cyan to green), and high error (yellow to red).} These results showcase how \uss{} models outperform iVR-GS composed models for unseen color and opacity mappings. For example, see "Broad Opacity" for chameleon, where the iVR-GS model incorrectly assigns opacity to large portions of the volume that are supposed to be transparent. }
    \label{fig:renders1}
    \vspace{-2em}
\end{figure*}

\subsection{Evaluation of Metrics}
\label{sec:benchmark}
For our evaluation, we create training and testing sets of $800\times800$ resolution images using PyVista's~\cite{sullivan2019pyvista} VTK backend~\cite{vtkBook}. We perform volume rendering with Blinn-Phong shading using a headlight from the camera for all images. We use 10 training opacity maps as described in Section~\ref{sec:training} for all datasets except RBL, where we use only 5 steps due to the dataset's sparsity. Multiple colormaps were experimented with, but since all led to similar results we choose to use only the Matplotlib~\cite{Hunter_matplotlib_2007} "viridis" colormap during training for simplicity. For every transfer function used for training or testing, we render 160 images where camera positions are sampled from a spherical orbit around the volume iterating through 16 azimuthal and 10 elevation angles. The camera zoom was selected as the minimum which shows the entire volume's bounding box in each image, and this design decision is motivated in Section~\ref{sec:ablation}. All testing images use camera views not seen during training, so that all results are for novel view synthesis. We create \uss{} models for each dataset by initializing Gaussians from 500,000 data points of the original volume. \new{While the number of Gaussians necessary for a model depends on the represented dataset's complexity, our densification and pruning methods allow the model to find this number during training, so an initial guess of 500,000 is sufficient.} As seen in Table~\ref{tab:evaluation_results}, models regularly end with drastically different numbers of Gaussians after training. All models are trained for 30,000 iterations with settings matching the defaults in 3DGS and iVR-GS wherever possible. Densification and pruning were enabled from iteration 500 until iteration 20,000. We explored using iVR-GS's two-stage training approach for Gaussians to learn geometry before scalar values and lighting effects, but found best performance with training all parameters from the first iteration. We use a learning rate of $0.00025$ for scalar values and $0.025$ for weights, along with a pruning threshold of $0.005$ for removing Gaussians with low weight. \new{These values were found by conducting a hyperparameter search on the Mito dataset, and choosing the values that led to the highest reconstruction quality to number of Gaussians ratio.}

We also create composed iVR-GS models for each dataset, following their approach of combining multiple basic sets of editable Gaussians trained from transfer functions with disjoint opacity bumps. For example, to train 10 basic sets, one will be trained using renderings where a transfer function left all values transparent except $[0,0.1]$, the next with all values transparent except $[0.1,0.2]$, and so on. We tested using flat opacity bumps for each basic set's transfer function, but found equal performance when training with the ramped opacity maps used by our method, so here we present results using ramped opacity maps for consistency with our method. Choosing an appropriate number of basic sets to train for each dataset is not trivial, as composing more sets improves reconstruction quality while worsening training time, memory footprint, and rendering performance significantly. The original paper~\cite{ivrgs_2025} does not provide an easy heuristic to decide where to make this tradeoff, so for our study we choose to compose 10 basic sets for all datasets except RBL, where we compose 5 due to its sparsity. As with our models, all training images use the "viridis" colormap. \new{We use the default settings for their method in all cases, and follow their two-stage training approach for a total of 40,000 training iterations per basic model.} The training time reported for iVR-GS models in Table~\ref{tab:evaluation_results} is the sum of training times for each of the basic sets, not including time to compose the sets together. The file size, number of Gaussians, and FPS reported are those of the composed models. All quality metrics are computed using renderings from the composed models. For testing of transfer functions, we apply colormaps by setting the palette color for each basic set to the colormaps' value at the corresponding portion of the function range, taken at the midpoint of the value range that set represents. We apply opacity mappings in the same way, assigning the opacity for each basic set according to the given opacity function.

\begin{figure*}[t]
    \centering
    \includegraphics[width=\linewidth]{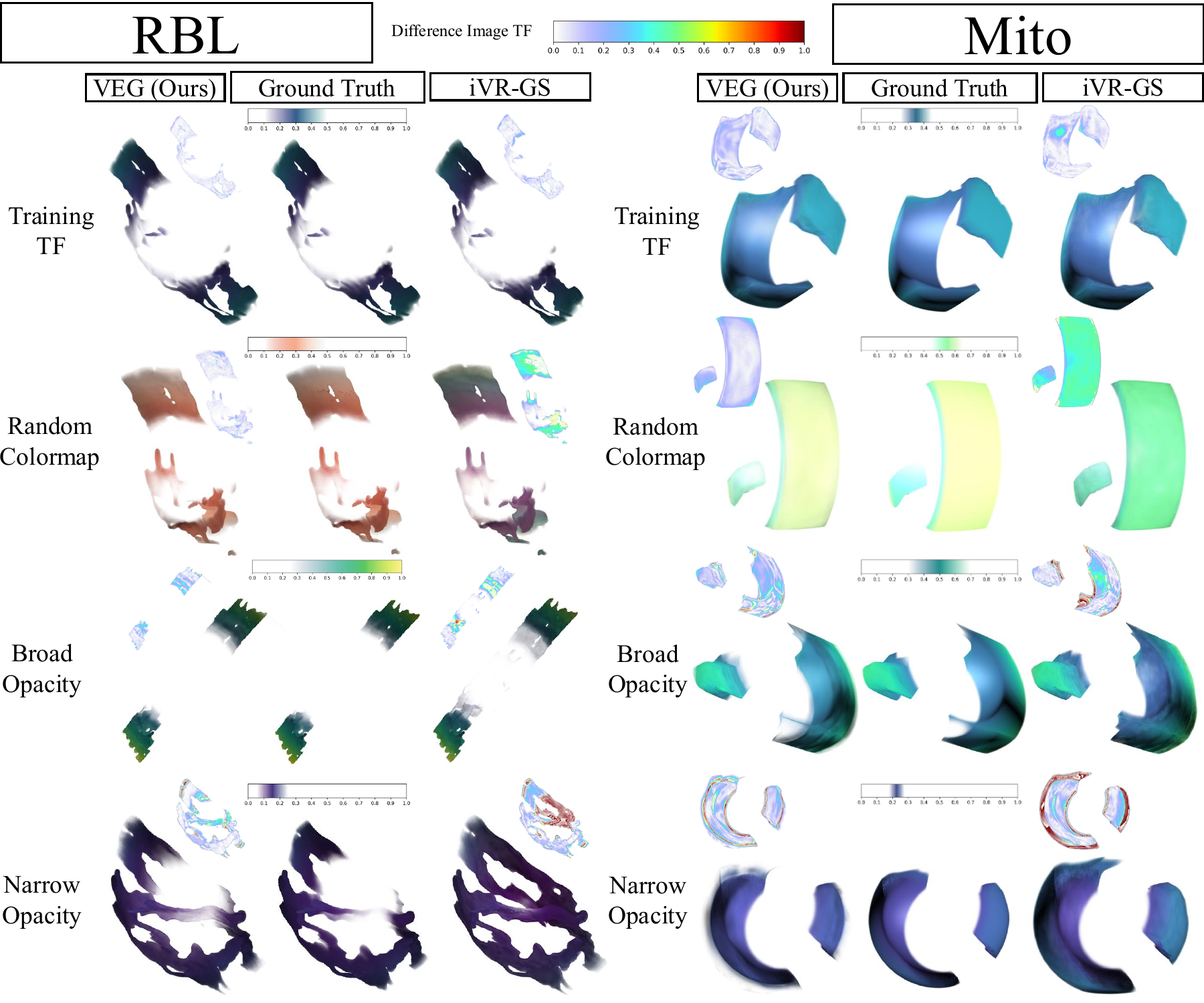}
    \caption{Qualitative comparison of our method with iVR-GS on the RBL and Mito datasets. We show results from each of the tests run in Table~\ref{tab:evaluation_results}, \new{along with difference images computed in the same way as Figure~\ref{fig:renders1}}. These results showcase how \uss{} models outperform iVR-GS composed models for unseen color and opacity mappings. See both of the "Random Colormap" sections, where iVR-GS models are incapable of applying the transfer function correctly, as it causes the basic sets to have palette colors perceptually different than during training. Also see "Broad Opacity" and "Narrow Opacity" sections, where the \uss{} models more closely follow the ground truth while the iVR-GS models incorrectly render more of the volume.}
    \label{fig:renders2}
    \vspace{-2em}
\end{figure*}

In order to measure the generalizability of both approaches for arbitrary transfer functions, we evaluate reconstruction quality on three use cases. First, performing novel view synthesis on the transfer functions seen during training. Next, rendering unseen views with colormaps that were unseen during training. We select five colormaps for variance with regards to color assigned to a specific part of the function range, along with the maps' popularity. Since all models were trained on Matplotlib "viridis", for testing we use "rainbow", "rainbow reversed", "cool to warm", "warm to cool", and "red blue yellow." Finally, we render unseen views with opacity mappings not seen during training. For this evaluation, we want to test each approach's ability to both infer volume renderings where large portions of the function range are visible, and surface-type renderings where only narrow portions of the function range are visible. To do this, we create sets of opacity maps where the slope of the opacity functions were either double the slope of the training transfer functions, leading to volume rendering-type images, or half the slope of the training transfer functions, leading to surface rendering-type images. We call these sets of opacity mappings "Broad Opacity" and "Narrow Opacity" in our results.

We present these results, along with training time, number of Gaussians in resulting models, file size, and rendering FPS in Table~\ref{tab:evaluation_results}. We find several conclusions from this study. First, we find that our method drastically improves on training time, number of Gaussians, and file size compared to iVR-GS, with average improvements of $6.76\times$, $16.28\times$, and $16.99\times$ respectively. These improvements can mostly be explained by our method directly training a single set of 3D Gaussians to represent a volume, while iVR-GS requires training multiple basic sets individually, then composing them to create their full volume model. Even so, the number of Gaussians in our trained \uss{} models is less than the number of Gaussians in a single basic set of the corresponding iVR-GS models. We find that \uss{} models are capable of offering a high-quality interactive rendering of the 4.2GB chameleon dataset while requiring only 2.7MB of memory and 1.6MB for storage. Interestingly, while we do find large frame rate speedups with our method, they are far from linearly correlated with the number of Gaussians, likely due to fixed costs per frame and higher GPU utilization with larger models. All models are able to achieve interactive framerates except for iVR-GS's Vortex model, showing the advantage of 3DGS-based methods for real-time rendering, even with a laptop GPU. 

\begin{table*}[t]
\centering
\resizebox{\textwidth}{!}{%
\begin{tabular}{@{}l|c|c|c|c|c|c|c|c|c|c|c@{}}
\toprule
\multirow{2}{*}{Ablation} & \multicolumn{2}{c|}{Training TF} & \multicolumn{2}{c|}{Unseen Colormaps} & \multicolumn{2}{c|}{Broad Opacity} & \multicolumn{2}{c|}{Narrow Opacity} & Training & \# of & File Size \\
& PSNR & SSIM & PSNR & SSIM & PSNR & SSIM & PSNR & SSIM & Time (min) & Gaussians & (MB)\\
\hline
Default & \textbf{36.36} & \textbf{0.99} & 33.75 & \textbf{0.99} & 28.00 & 0.97 & 28.08 & 0.98 & 17.1 & 49,969 & 1.1\\
Random Init & 35.53 & \textbf{0.99} & \textbf{33.91} & 0.98 & 27.76 & 0.97 & 28.19 & 0.98 & 20.5 & 40,906 & \textbf{0.9}\\
No Weights & 33.65 & \textbf{0.99} & 33.72 & \textbf{0.99} & 24.76 & 0.95 & 25.92 & 0.97 & 37.8 & 563,223 & 13.3\\
Linear Opac & 18.28 & 0.93 & 17.45 & 0.93 & 17.37 & 0.92 & 18.66 & 0.93 & 27.0 & 70,492 & 1.6\\
5 Opac Steps & 26.23 & 0.97 & 26.92 & 0.97 & \textbf{36.08} & \textbf{0.99} & 25.82 & 0.97 & 18.2 & 88,678 & 2.1\\
20 Opac Steps & 29.28 & 0.98 & 28.18 & 0.97 & 28.18 & 0.97 & \textbf{34.74} & \textbf{0.99} & \textbf{12.4} & \textbf{39,001} & \textbf{0.9}\\
\bottomrule
\end{tabular}
}
\caption{\new{Evaluation metrics for novel view synthesis with different ablations of our method, evaluated on the Mito dataset. Best results in bold.}}
\label{tab:ablation_results}
\vspace{-1em}
\end{table*}

Next, we find that while iVR-GS outperforms our method for novel view synthesis using the transfer functions that the models were trained on, our method can more accurately infer novel views for transfer functions unseen during training. When generalizing to unseen colormaps, \uss{} models consistently outperform  iVR-GS models, with an average increase of 5.64 PSNR. With regards to broad opacity maps, \uss{} models have an average increase of 4.24 PSNR. Finally, for narrow opacity maps, while iVR-GS actually outperforms \uss{} on Chameleon, \uss{} models still produce an average increase of 3.02 PSNR. \new{These results show that with our training of scalar field values directly, we are able to create a transfer function-agnostic model in only 30,000 iterations of training, while iVR-GS's color storing approach requires a total of 400,000 (200,000 for RBL) iterations of training while creating models with worse accuracy for transfer functions unseen during training. }

We present a qualitative evaluation of our method in Figures \ref{fig:renders1} and \ref{fig:renders2}, for structured and unstructured datasets respectively. We include a rendering for both our method and iVR-GS for every dataset, for each of the reconstruction quality tests assessed in Table~\ref{tab:evaluation_results}. Ground truth renderings are given as well, along with images showing the pixel-wise perceivable difference between each reconstructed rendering and the ground truth. By analyzing these images, it's possible to see what causes the difference in reconstruction quality between methods for each type of test. Starting with the tests on training TFs, iVR-GS models outperform \uss{} for two reasons. First, by training color directly instead of optimizing scalar values through a transfer function, iVR-GS models can more accurately reconstruct the exact rendered colors of the ground truth. Second, iVR-GS models contain many more Gaussians than \uss{} models, so they are better able to approximate fine-grained volume geometry for the collection of surfaces that are seen during training. For the random colormap tests, iVR-GS's method of editing Gaussian color by changing the palette color for each basic set does not generalize well when the color being applied has high perceptual difference to the original palette color. In contrast, \uss{} models handle recoloring through direct application of transfer functions. In both of the opacity mapping tests, the main reason for higher reconstruction quality with \uss{} models appears to be avoiding assigning opacity to areas that should not be rendered, illustrated by the Chameleon results for "Broad Opacity" in Figure~\ref{fig:renders1}. Storing scalar values allows \uss{} models to cull Gaussians more precisely where the opacity function determines, rather than iVR-GS which can only control opacity set-wide.

\begin{figure*}[t]
  \centering
  \includegraphics[width=\linewidth]{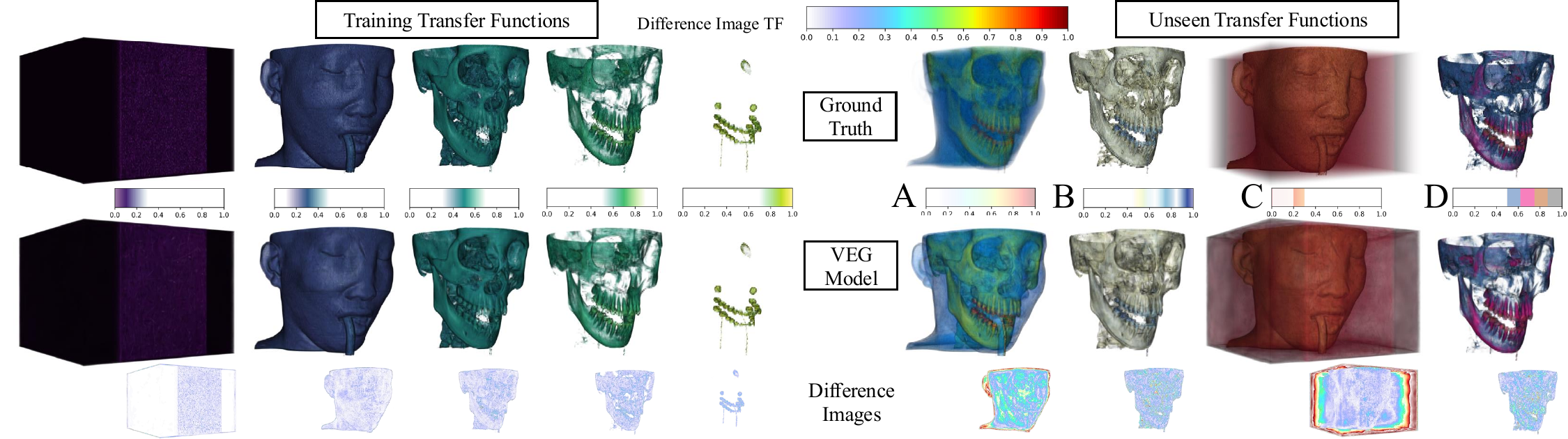}
  \caption{Qualitative showcase of novel view synthesis for a \uss{} model on complex unseen transfer functions. We train a model for the Paraview head dataset using the five transfer functions on the left, and test with those and the four functions on the right (A-D), where the opacity and color maps were unseen during training. The top, middle, and bottom rows of images show ground truth, inference, and difference images (as in Figure~\ref{fig:renders1}). }
  \label{fig:teaser}
  \vspace{-1em}
\end{figure*}

\subsection{Ablation Study}
\label{sec:ablation}
\new{In this section, we conduct several tests for validation of design decisions made for \uss{}. Specifically, we test the effect of our initialization strategy, our use of additional weight parameters per Gaussian, our opacity-guided training strategy, the generalizability of training with different numbers of opacity maps, and optimal zoom level for training image selection. To do this, we train models on the Mito dataset while changing various aspects of our algorithm. Unless otherwise stated, training and testing follow the procedure from Section~\ref{sec:benchmark} exactly. We present our results in Table~\ref{tab:ablation_results}. The "Default" row represents usage of the default parameters and method of our algorithm, and the metrics are taken directly from Table~\ref{tab:evaluation_results}}.

\textbf{Initialization} \new{To test our volume initialization strategy, we compare our default initialization against a model trained starting from 500k data points with random positions and values. These results are given in the "Random Init" row of Table~\ref{tab:ablation_results}. Surprisingly, we find that the performance of this model is around equal to the default model in all cases except the training TF, where the default's quality is higher. This shows that our method is robust to random initialization, likely due to the densification and pruning strategies that allow creating and destroying new geometry as needed.}

\textbf{Weight Parameter Training} \new{As mentioned in Section~\ref{sec:rendering}, our \uss{} models include a trainable weight parameter per Gaussian that is multiplied with the transfer function-determined opacity when rendering. This allows for more direct model optimization, along with the ability to remove ineffective Gaussians as their weight parameters are pushed below our pruning threshold during training. To test this design decision, we present results from training a \uss{} model without the use of this parameter in Table~\ref{tab:ablation_results}, labeled "No Weights". As can be seen, the reconstruction quality of this model dips significantly from the default model. In addition, because there is no way to prune based on the weight threshold, the model can only grow from its starting 500,000 Gaussians, ending with 563,223 compared to the default model's 49,969.}

\textbf{Multiple Opacity Maps} \new{An important contribution of our work is our opacity-guided training, in which we use chosen opacity maps to ensure that the entire function range of a volume is accurately represented. To test this, we compare training with our default 10 steps of opacity maps against naive training with a single linear opacity map. These results are given in the "Linear Opacity Map" row of Table~\ref{tab:ablation_results}. This model achieved low quality for all other sets of opacity maps, illustrating that the model did not learn the dataset's full function range.}

\new{In addition, we experiment with the number of opacity maps necessary for accurate volume reconstruction. We conduct training using our multiple opacity map method from Section~\ref{sec:training} with number of steps 5 ("5 Opac Steps"), 10 ("Default"), and 20 ("20 Opac Steps"), then evaluate how well each model generalizes to each of the other sets of opacity maps. We present the results in Table~\ref{tab:ablation_results}, with "Training TF" testing the opacity maps from the set with 10 steps, "Broad Opacity" testing from the set with 5 steps, and "Narrow Opacity" testing from the set with 20 steps. As expected, each model outperformed the others when tested on the set of opacity maps it was trained on. We also find, as the number of training opacity maps increases, the model seems to perform better on opacity maps not seen during training. This is likely because wider opacity maps can prevent the model from learning features for regions of the volume that are overlapping for that map's portion of the function range. Interestingly, the training time and number of Gaussians in the resulting models correlates inversely with the number of opacity maps used for training, likely because narrower opacity maps lead to sparser training images. }

\begin{table}[t]
\centering
\resizebox{\columnwidth}{!}{%
\begin{tabular}{@{}l|c|c|c|c|c|c|c@{}}
\toprule
\multirow{2}{*}{Training Set} & \multicolumn{2}{c|}{Default} & \multicolumn{2}{c|}{Zoomed In} & Training & \# of \\
& PSNR & SSIM & PSNR & SSIM & Time (min) & Gaussians \\
\hline
Default & 36.36 & 0.99 & 33.34 & 0.98 & 17.1 & 49,969 \\
Zoom & 34.87 & 0.99 & 33.43 & 0.98 & 22.7 & 119,821 \\
\bottomrule
\end{tabular}
}
\caption{\new{Evaluation metrics for novel view synthesis for default and zoomed in testing sets of our method, evaluated on the Mito dataset. We compare our model trained on the default training set and training set including zoomed images. The default model outperforms on the default test set, while maintaining near equal quality on the zoomed in test set. }}
\label{tab:zoom}
\vspace{-1.5em}
\end{table}

\textbf{Optimal Zoom Level} \new{We also validate our selection of zoom level for training images. We create a set of test images identical to our Training TF testing set from Section~\ref{sec:benchmark} for the Mito dataset, but with varying zoom levels at either 1 (default), 2, or 3$\times$. We name this testing set "Zoomed In," and report the performance of our default model on this and the default testing set in Table~\ref{tab:zoom}. Additionally, we train another model "Zoom," which is identical to the default except that its training image set also included images at 2 and 3$\times$ zoom, and report its results on both testing sets. We find that, while our model does perform worse when rendering zoomed in views, including similarly zoomed views in the training set does not improve model quality, validating our camera sampling strategy for training images. In fact, the "Zoom" model shows noticeably worse performance on the default views, while using more than 2$\times$ the number of Gaussians as the default model.}

\subsection{Complex TF Showcase}
\label{sec:head}
The final evaluation we present is to test our method on more complex transfer functions than those in Section~\ref{sec:benchmark}. To do this, we use a Paraview example dataset of a CT scan of a boy's head made up of a structured volume of 256x256x94 cells. We choose this dataset as it consists of easily highlightable anatomical features, allowing for examples such as the opaque skull rendered underneath semi-transparent skin. We train a single model for this dataset in the same way as previously, using 5 opacity maps. We then manually select 4 transfer functions for testing, where the opacity and color maps were both unseen during training. From A to D in Figure~\ref{fig:teaser}, these were selected for demonstrating a linear opacity map over the full function range, multiple narrow opacity bumps, a semi-transparent opacity bump along with a high opacity bump, and a constant opacity as well as a qualitative colormap. Results of the \uss{} model for novel view synthesis on the training and unseen transfer functions are given in Figure~\ref{fig:teaser}. While the model performs more poorly on our custom transfer functions (especially in overfilling semi-transparent areas), these results show that \uss{} models can perform high-quality novel view synthesis even with arbitrary transfer functions disparate from those seen during training. 
% \section{Results}
% We now discuss key insights and trade-offs that were observed from our experiments, highlighting the practical strengths and limitations of \uss{}.
\section{Conclusion}
\label{sec:conclusion}

In this paper, we introduced \textit{\us{} (\uss{})}, a novel 3D Gaussian splatting (3DGS) approach tailored specifically for structured and unstructured scientific volume rendering. \uss{} departs from conventional 3DGS techniques by removing color and opacity encodings and instead directly encoding scalar valued fields from volume datasets. This decouples the visual appearance from data representation, enabling lightweight, transfer function-agnostic models suitable for interactive visualization. Our evaluation demonstrates that \uss{} can achieve high reconstruction quality with robust generalization to previously unseen transfer functions, while maintaining tiny storage sizes, requiring less than 5 megabytes for all datasets studied.

For future work, we will explore support for time-varying volumetric data and distributed training schemes in order to improve scalability for very large datasets and enable training in HPC contexts. Additionally, we are interested in investigating the possibility of \uss{} to fully reconstruct volumes, rather than only render images. Future work will look into creating training pipelines for compact, high-quality \uss{} models to use as compressed representations for volume sampling directly. These could then be rendered through 3D Gaussian raytracing, similarly to recent work~\cite{loccoz20243dgrt}.

\section{Acknowledgments}
This work was partly funded by NSF Collaborative Research Awards 2401274, 2221812, and 2106461, and NSF PPoSS Planning and Large awards 2217036 and 2316157.

% We have several plans for extending and refining this work in the future. First, our current pipeline relies on PyVista for rendering ground-truth images of datasets. PyVista was selected over native VTK or a custom unstructured volume renderer to accelerate early development. However, PyVista struggles with performance, particularly for large unstructured volumes, and lags behind the latest versions of VTK. Transitioning directly to VTK or developing a custom renderer optimized for unstructured data would substantially improve the robustness of our training process.

% Second, while existing 3DGS renderers are incredibly full-featured, our approach requires post-processing to convert trained models into a compatible format for them. Because of this, existing interactive renderers are not capable of showcasing the full benefits of \uss{}. Developing a custom interactive renderer specifically tailored to \uss{}, capable of applying custom transfer functions dynamically, would greatly enhance the usability and effectiveness of our method for real-time scientific data exploration.

% Finally, integrating recent advancements from the broader 3DGS community could yield additional performance improvements. 

%% if specified like this the section will be omitted in review mode
% \acknowledgments{%
% 	The authors wish to thank A, B, and C.
%   This work was supported in part by a grant from XYZ (\# 12345-67890).%
% }

\bibliographystyle{abbrv-doi-hyperref}

\bibliography{template}

\end{document}